\documentstyle[preprint,aps]{revtex}

\input psfig

\tightenlines

\begin{document}

\draft

\title{\bf STUDIES OF A 2-CHAIN SPIN LADDER\\
WITH FRUSTRATING SECOND NEIGHBOR INTERACTIONS}
\author{Zheng Weihong\cite{byline1}, V. Kotov\cite{byline2}, J. Oitmaa\cite{byline3}} 
\address{School of Physics,                                              
The University of New South Wales,                                   
Sydney, NSW 2052, Australia.}                      

\date{Oct. 31, 1997}
\date{\today}

\maketitle 

\begin{abstract}
The Heisenberg model on a 2-chain spin-${1\over 2}$ ladder 
with frustrating second neighbor 
interactions is studied by using series expansions about
the Ising and dimer limits, numerical diagonalization, and
 by Abelian bosonization analysis. 
The phase diagram is determined,
and pair correlations and the complete dispersion relations for the triplet 
spin-wave excitations are also computed.
\end{abstract}                                                              
\pacs{PACS Indices: 75.10.-b., 75.10J., 75.40.Gb  }


\narrowtext
\section{INTRODUCTION}
Heisenberg spin ladders have been  the subject
of intense theoretical and experimental research in recent years.
It is by now well established that  single chain 
Heisenberg antiferromagnets
with integer spin have a gap in the excitation spectrum, whereas 
those with half-integer spin  have gapless excitations.
The former have a finite correlation length, while
for the latter it is infinite with the spin-spin
correlation function decaying as a power law.
For $S={1\over 2}$ Heisenberg spin ladders\cite{dag95,gre96,ladder}, 
the systems with an even number of legs have an 
energy gap, short range correlations and a ``spin liquid" ground
state. On the other hand, the systems with odd number of legs have 
gapless excitations, quasi long range order, and a power-law falloff of
spin-spin correlations, similar to single chains. 

In this paper, we study the Heisenberg model on a 2-chain ladder with 
second neighbor interactions,  
via Ising and dimer expansions, diagonalization of finite systems,
and  Abelian bosonization.
All of the work is at $T=0$. The motivation for studying such a system
is twofold. Firstly we wish to explore the effect of frustration
on the properties of 2-leg $s={1\over 2}$ ladders. Weak frustration is
not expected to change the physics of the gapped system qualitatively.
However strong frustration may change the nature of the ground state
at some critical coupling ratio, corresponding to a $T=0$
phase transition. Furthermore in other systems, such as the
$J_1-J_2$ chain, frustration itself leads to the creation of a gapped
dimer phase, and thus in the present case there is the possibility
of observing the competition between two independent gap yielding
perturbations. The second reason for studying such a generalized ladder
system is the possibility that in real 2-leg ladder materials
significant second neighbour interactions will be present.

We write the Hamiltonian of our system as
\begin{eqnarray}
H(J_{{}_{/\!/}},J_{\perp}, J_2) = && \sum_{i} [ J_{{}_{/\!/}} 
   ({\bf S}_{1,i} \cdot {\bf S}_{1,i+1}
 + {\bf S}_{2,i} \cdot {\bf S}_{2,i+1})
 + J_{\perp}  {\bf S}_{1,i}\cdot {\bf S}_{2,i}  \nonumber \\
 && + J_2 ( {\bf S}_{1,i}\cdot {\bf S}_{2,i+1} 
 + {\bf S}_{2,i}\cdot {\bf S}_{1,i+1} ) ]
\label{H}
\end{eqnarray}
where ${\bf S}_{l,i}$ denotes the $S=1/2$ spin at 
site $i$ of the $l$th chain. $J_{{}_{/\!/}}$ is the 
interaction between nearest neighbor spins 
along the chain, $J_{\perp}$ is the interactions between 
nearest neighbor spins along the rungs, $J_2$ is the interactions between
the second neighbor spins. 
This is shown in Figure 1(a).
We denote the ratio of couplings as
$y_1\equiv J_{\perp}/J_{{}_{/\!/}} $ and $y_2\equiv J_2/J_{{}_{/\!/}} $.
In the present paper 
all couplings are taken to be antiferromagnetic
(that is, $J_{{}_{/\!/}},~ J_{\perp}, ~J_2 >0$).

The system has an interesting symmetry property:
If one exchanges the couplings $J_{{}_{/\!/}}$ and $J_2$, one
can recover the original Hamiltonian by exchanging two spins along
the rungs at even sites, that is, the Hamiltonian
will be invariant under exchanging the couplings $J_{{}_{/\!/}}$ 
and $J_2$:
\begin{equation}
H(J_{{}_{/\!/}}, J_{\perp}, J_2 ) = 
H(J_2,  J_{\perp} , J_{{}_{/\!/}} ) \label{Hsym}
\end{equation}
Therefore we only need to study the case of $y_2\leq 1$, and 
the system with
$y_2> 1$ can be mapped into system with $y_2\leq 1$ through the identity
in Eq. (\ref{Hsym}).

There has been some previous work on this system. Of course for $J_2=0$
we recover the usual 2-leg ladders\cite{dag95,gre96,ladder}, which has 
gapless excitations only when $J_{\perp}=0$.
The case $y_2=1$ has  special properties.
Bose and Gayen\cite{bose} first pointed out that this system
has an  exact dimer state: a state in which every pair of spins along 
the rungs form a singlet,
and this perfect dimer state is the ground state for large enough
values of $y_1$. 
Xian\cite{xian} performed a systematic study of this system via
a microscopic approach based on a proper set of
composite operators, and found that the Hamiltonian 
consists of two commuting parts:
\begin{equation}
H = \sum_{i} J_{\perp}  {\bf S}_{1,i}\cdot {\bf S}_{2,i} 
+ J_{{}_{/\!/}} {\bf P}_{i}\cdot {\bf P}_{i+1} 
\end{equation}
where ${\bf P}_i$ represent the effective spin-1 operator at site $i$.
The first part, $J_{\perp}  {\bf S}_{1,i}\cdot {\bf S}_{2,i}$, 
is trivial, with no interaction between different rungs and with a gap
$J_{\perp}$ between the singlet and triplet states.
The second part, $J_{{}_{/\!/}} {\bf P}_{i}\cdot {\bf P}_{i+1} $,
is similar to the spin-1 Heisenberg chain with each rung in the
ladder corresponding to each site in
the spin-1 chain.
Because the two part of $H$ commute it follows that there is an
eigenstate of $H$ with eigenvalue:
\begin{equation}
E_g = (J_{{}_{/\!/}} e_0  + J_{\perp}/4 ) N/2
\end{equation}
where $e_0=-1.40148403897(4)$  is the ground-state energy per site of the
spin-1 Heisenberg chain\cite{dnrg}.
The eigenvalue for the state with singlets on all of the rungs is
\begin{equation}
E_g = (-3 J_{\perp}/4 ) N/2
\end{equation}
Therefore one can easily 
determine the transition point $y_{c}= 1.40148403897(4)$.
At this crossing point the singlet-singlet gap is zero.
Kitatani and Oguchi\cite{kit96} had independently considered the
existence of a state of singlet dimers for this system for $y_2=1$,
and had obtained the same results as above.

This paper studies the general case where exact results are not known.
In Section 2 we report results obtained by Ising and dimer expansions
at $T=0$. We compute the ground state energy and the complete spin-wave
excitation spectra, as well as obtaining more accurate estimates
for various quantities for the simple case $J_2=0$. We also
carry out finite lattice diagonalization studies for systems of
16 and 24 spins, paying particular attention to the
behaviour of spin-spin correlations. Section 3 describes
an analytic approach, based on Abelian bosonization.
When comparisons are possible the various methods all 
provide a consistent picture. Finally in Section 4 we
provide a summary and draw some conclusions.

\section{Numerical Results}
We report here the results of numerical studies, using both series
expansions and diagonalizations. These methods provide results over
the whole phase diagram, and give the most detailed picture of 
the behaviour of this system for general values of the couplings.

Our previous series work on the ladder system\cite{ladder} used
expansions about both the Ising limit and about a fully
dimerized state, and we refer to that paper and references
therein for technical details.

\subsection{Ising expansions}
To construct an expansion about the Ising limit for this system, 
one has to introduce an anisotropy parameter $x$, 
and write the Hamiltonian as:
\begin{equation}
H = H_0 + x V  \label{Hising}
\end{equation}
where 
\begin{mathletters}
\begin{eqnarray}
H_0 &= & \sum_{i} [ J_{{}_{/\!/}} 
( S_{1,i}^z S_{1,i+1}^z + S_{2,i}^z S_{2,i+1}^z )
+  J_{\perp}   S_{1,i}^z S_{2,i}^z +
 J_2  ( S_{1,i}^z S_{2,i+1}^z + S_{2,i}^z S_{1,i+1}^z ) ] \label{eqh0} \\
V &= &  \sum_{i} [ J_{{}_{/\!/}}  ( S_{1,i}^x S_{1,i+1}^x 
+ S_{1,i}^y S_{1,i+1}^y 
+ S_{2,i}^x S_{2,i+1}^x + S_{2,i}^y S_{2,i+1}^y )  + 
J_{\perp} ( S_{1,i}^x S_{2,i}^x + S_{1,i}^y S_{2,i}^y ) \nonumber \\
&& + J_2  ( S_{1,i}^x S_{2,i+1}^x + S_{1,i}^y S_{2,i+1}^y 
             + S_{2,i}^x S_{1,i+1}^x + S_{2,i}^y S_{1,i+1}^y ) ]
\end{eqnarray}
\end{mathletters}
The limits $x=0$ and $x=1$ correspond respectively to the Ising  and
isotropic Heisenberg limit, the latter being the case of
primary interest.

Since $H_0$ is taken as the unperturbed Hamiltonian we
need to identify the ground states. There are three
of these, shown in Figure 1(b). We refer to these as 
the N\'{e}el state (N), the ferromagnetic chain state (F)
and the ferromagnetic rung state (R). Their regions of
stability are indicated in Figure 2. The line
$y_2=1$, $y_1\geq 2$ is a boundary between
N and F states. The operator $V$ is treated as
a perturbation: it flips pairs of spins on neighbouring sites.
The quantum fluctuations represented by $V$ will of 
course result in much more complex ground states. Since the
system is effectively 1-dimensional no true long range
order can exist, even at $T=0$.

As in our earlier paper\cite{ladder}, to overcome a 
possible singularity at $x<1$, 
and to get a
better convergent series in the Heisenberg limit, 
we add to the Hamiltonian  the following staggered 
field term 
\begin{equation}
\Delta H = t (1-x) \sum_{i,l} (-1)^{i+l} S_{l,i}^z
\end{equation}
for the expansion about the Ne\'{e}l state, or the following  
field term 
\begin{equation}
\Delta H = t (1-x) \sum_{i,l} (-1)^{i} S_{l,i}^z
\end{equation}
for the expansion about the ferromagnetic rung state.
$\Delta H$ vanishes at the isotropic limit $x=1$. 
We adjust the coefficient $t$ to 
get the most smooth terms in the series, 
with a typical value being $t=2$.

The series expansion method 
has been previously described in several  
articles\cite{he90,gel90,gelmk}, and will not be repeated here.
We have developed Ising expansions about both the
Ne\'{e}l  and ferromagnetic rung states for 
the ground state energy per site $E_0/N$ and the triplet 
spin-wave excitation spectrum 
for different ratios of 
couplings $y_1$ and $y_2$ and (simultaneously) 
for several values of $t$ up to order $x^{13}$.
The resulting series are available on request. 
These series have been analyzed by using  integrated 
first-order inhomogeneous
differential approximants and Pad\'{e} approximants\cite{gut}.
We will discuss these results later in this section.

\subsection{Dimer expansions}
In the limit that the exchange coupling along the rungs $J_{\perp}$
is much larger than the couplings $J_{{}_{/\!/}}$ and $J_2$,
the rungs interact only weakly with each other, and the 
dominant configuration in the ground state is the product
state with the spins on each rung forming a spin singlet. 
The Hamiltonian in Eq. (\ref{H}) can then be rewritten as,
\begin{equation}
H/J_{\perp} =  H_0 + (1/y_1) V  \label{Hdimmer}
\end{equation}
where
\begin{eqnarray}
H_0 &=& \sum_{i} {\bf S}_{1,i} \cdot {\bf S}_{2,i} \nonumber \\  && \\
V &=& \sum_{i}[ ( {\bf S}_{1,i} \cdot {\bf S}_{1,i+1} 
  + {\bf S}_{2,i} \cdot {\bf S}_{2,i+1} ) 
  + y_2 ( {\bf S}_{1,i}\cdot {\bf S}_{2,i+1} + 
  {\bf S}_{2,i}\cdot {\bf S}_{1,i+1} ) ]
\nonumber 
\end{eqnarray}
We can obtain an expansion in $(1/y_1)$ by
treating the operator $H_0$ as the unperturbed
Hamiltonian, and
the operator $V$ as a perturbation.

We have carried out the dimer expansions for the ground state 
energy up to order $(1/y_1)^{23}$ 
and for the triplet excitation spectrum
up to order $(1/y_1)^{13}$ for different values of $y_2$.
The resulting series for some particular values of $y_2$ are 
listed in Table I. The rest of the series are 
available  on request. In our previous paper\cite{ladder},
the series for the case $y_2=0$ were
computed up to order $(1/y_1)^9$ for
the ground state energy and up to order $(1/y_1)^8$
for the triplet excitation spectrum.
Our present results agree with and considerably 
extend these previous results.
Again, we use integrated first-order inhomogeneous
differential approximants and Pad\'e approximants\cite{gut} 
to extrapolate the series.

The ground state energy per site $E_0/N$ is shown in Figure 3,
where curves for $y_2=0,0.2,0.4,0.6,0.8,1$ as shown as
functions of $y_1=J_{\perp}/J_{{}_{/\!/}}$. 
The curves (connecting the full point symbols)
emanating  from the left side (small $y_1$) are obtained from 
Ising expansion about the R state. Those emanating from
the right side are from the dimer expansion.
The estimates from the Ising expansion about the N state agree
very well with these and are not shown separately
(except for the case of $y_2=0$ which are shown
by open point symbols with error bars).
These curves, for any given $y_2$, cross at 
a transition point $y_{1c}$. This corresponds to
a first order ground state phase transition, resulting from
a level crossing. The numerical estimate for
$y_2=1$ of $y_{1c}=1.40$ is in good agreement with the
exact result of Xian\cite{xian} and Kitatani and Oguchi\cite{kit96}
discussed above. The locus of the transition points
is shown in Figure 2, and represents a ``gapless line''
where the gap between the two lowest energies, both
of which are singlets, vanishes. Everywhere else the
ground state is a non-degenerate singlet.
However this is not a gapless line
in the usual sense of a vanishing gap for elementary
excitations.

There are two branches of triplet spin excitations which
we denote $\epsilon_g(k)$, $\epsilon_u(k)$ corresponding to
symmetric and antisymmetric states with respect to
interchange of the two chains. $k$ is the wavenumber along the
chain direction.
On the left side of the gapless line in Fig. 2, which can be 
called a ``Haldane'' phase, these two branches are 
independent and each has the appearance of a simple
cosine dispersion curve, symmetric about $k=\pi/2$
and having a finite gap at the minimum energy points.
However on the right side of the gapless line, the
``dimerized region'', the branches are no longer independent,
being related by
\begin{equation}
\epsilon_g (k)=\epsilon_u (\pi-k)
\end{equation}
and having a complex form, without the symmetry about
$k=\pi/2$ found in the Haldane region.
This relation can be understood as follows.
Consider an initial down spin at site $j$ flipped to
create an excitation. To form a Bloch state this excitation 
will couple to other downs spins. In the Ne\'el or small
$y_2$ region this involves even sites on the upper chain
and odd sites on the lower chain leading to a shift of $\pi$ in
the wavenumber.
%
%
The dispersion curves are shown in Figs. 4-5 for 
$y_1=1$ for particular values of $y_2$.
The triplet gap appears to be nonzero throughout 
the phase diagram, except at $y_1=y_2=0$.

Since the dimer expansion carried  out here is much 
longer than our previous calculations\cite{ladder},
we can make more  accurate estimates for the normal 2-leg
ladder without frustration ($J_2=0$).
The ground-state energy is estimated to be
\begin{equation}
E_0/(NJ_{{}_{/\!/}}) =-0.578043(2) \quad {\rm at} \quad J_{{}_{/\!/}}=J_{\perp}~.
\end{equation}
In Fig. 6 we show the dependence of the triplet excitation gap
$\epsilon_u(\pi)$ on $J_{\perp}/(J_{\perp}+J_{{}_{/\!/}})$, in particular,
they are estimated to be
\begin{eqnarray}
\epsilon_u(\pi)/J_{{}_{/\!/}}&&=0.5028(8) \quad {\rm at} \quad J_{{}_{/\!/}}=J_{\perp}~;  \\
\epsilon_u(\pi)/J_{\perp}&&=0.405(15) \quad {\rm at} \quad J_{{}_{/\!/}} \gg J_{\perp}~.
\end{eqnarray}
These results agree very well with the recent Quantum Monte Carlo (QMC)
results of Frischmuth {\it et al.}\cite{fri96} and  Greven, Birgeneau, 
and Wiese\cite{gre96}.

\subsection{Exact diagonalizations}
In order to obtain a more complete picture of the energy spectrum,
the correct assignment of spin quantum number
$S$ to different levels, and the variation of spin-spin correlations
throughout the phase diagram we have carried out exact Lanczos
diagonalizations for ladders with $N=16$ and $N=24$ spins.
The finite lattice corrections appear to be quite small and the
results are believed to be representative of the thermodynamic
limit.

In Fig. 7 we show the energies of low lying energies 
for $N=24$, for a scan 
through the phase diagram (Fig.2)  at $y_2=0.8$.
The solid lines represent singlets ($S=0$), while
triplet levels ($S=1$) are shown as dashed lines.
The existence of the transition point discussed above, the
finite singlet-triplet gap, and other level crossings can be seen. 
There is a region around the transition point where the lowest
triplet lies above the lowest two singlets. 
The location of the transition point, while less accurately
resolvable, is totally consistent with the
series results given above. Fig. 8 we show some pair correlations, for the same
scan through the phase diagram at
$y_2=0.8$.
The correlations show large discontinuities at the
transition point $y_1\simeq 1.24$. The sign 
of various correlations is consistent with ferromagnetic
rung (R) type order for $y_1< 1.24$ and N\'{e}el type order for
$y_1> 1.24$, in accordance with the classical ground states
in Figure 1. We note also that further neighbour correlations become
very small for increasing $y_1$, consistent with a dimerized
phase.


\section{Weak Coupling Analysis}
In this section we present analysis, based on Abelian  bosonization,
which can   determine the phase boundary between the 
Ferromagnetic rung (Haldane-type)  and N\'{e}el 
ground states.
Our considerations are valid in the limit of small
$J_{\perp}$ and $J_2$. Since the procedure is well described in the
literature \cite{Luther,Schulz,Millis}, we will give  only the basic steps here.

The spin operators for each chain  are transformed, by using  the Jordan-Wigner
transformation, into a system of spinless fermions
$a_{n}$(chain 1) and $b_{n}$(chain 2)   at half filling.
Next, since we are interested in the low-energy properties of the
model, we pass to a continuum description, which was developed for a single
chain by Luther and Peschel \cite{Luther}. The  spectrum of the
Jordan-Wigner fermions is linearized in the vicinity of the
two Fermi points $\pm k_{F} = \pm \pi/2$:

\begin{equation}
a_{n} = \sqrt{a}[e^{ik_{F}n} \psi_{1R}(x=na) + e^{-ik_{F}n} \psi_{1L}(x=na)],
\end{equation}   
where $a$ is the lattice spacing, and $\psi_{1R}$, 
$\psi_{1L}$ are slowly varying on the scale
of the lattice. An analogous transformation is applied to the second chain
(with corresponding fields $\psi_{2R}$, $\psi_{2L}$). 
In order to write the ladder Hamiltonian (1) in bosonic form, we make use of the
Abelian bosonization rules:        

\begin{equation}
\psi_{1R,1L} = \frac{1}{\sqrt{2\pi \alpha}} \mbox{exp}[\pm i \sqrt{4\pi} 
\phi_{1R,1L}(x)],
\end{equation}
where $\alpha^{-1} \sim a^{-1}$  is a large momentum cutoff, and:

\begin{equation}
\phi_{1R,1L}(x) = \frac{1}{2} \left[ \phi_{1}(x) \pm \theta_{1}(x) \right].
\end{equation}
Here $\theta_{1}(x)$ is defined as the field, dual
to $\phi_{1}(x)$, i.e. $\partial_{x}\theta_{1}(x) = 
\Pi_{1}(x)$, and $\Pi_{1}(x)$ is the momentum field, conjugate to $\phi_{1}(x)$.
Similar equations describe the second chain.
Using the above formulas the interchain interactions can be bosonized,
with the result:    

\begin{equation}
H = \sum_{s=+,-}\frac{v_{s}}{2} \int dx \left[\frac{1}{K_{s}} 
(\partial_{x}\phi_{s})^{2} + K_{s}(\partial_{x}\theta_{s})^{2}
\right] + \tilde{H}, \label{eq19}
\end{equation}
\begin{equation}
\tilde{H} = \frac{C}{\alpha^{2}} \int dx \left[ g_{1}
 \mbox{cos}(\sqrt{8\pi} \phi_{+}) + g_{2}\mbox{cos}(\sqrt{8\pi} \phi_{-})+
 2g_{3}\mbox{cos}(\sqrt{2\pi} \theta_{-}) \right]. \label{eq20}
\end{equation}
In the above equations, the symmetric and antisymmetric combinations
of the fields are introduced via $\sqrt{2}\phi_{\pm}= \phi_{1} \pm \phi_{2}$,
and similarly for the dual fields. The (bare) values of the
coupling constants in (\ref{eq20}) are $g_{i} = y_{1} -2y_{2} , i=1,2,3$, and
$C$ is a cutoff independent constant. The Luttinger liquid   
parameters in (\ref{eq19}) depend on the bare couplings:

\begin{equation}
v_{\pm} = 1 + \frac{\delta}{\pi} \pm \frac{y_{1}+2y_{2}}{2\pi}
+ \mbox{higher order},\label{eq21}
\end{equation}
\begin{equation}
K_{\pm} = 1 - 2 \frac{\delta}{\pi} \mp \frac{y_{1}+2y_{2}}{2\pi}
+ \mbox{higher order}.\label{eq22}
\end{equation}
Without the $y_{2}$ terms our equations are similar to  the ones
obtained by Strong and Millis  for the simple ladder \cite{Millis}.
In order to emphasize that the second terms in (\ref{eq21}) and (\ref{eq22}) come from
the Ising interactions in the two chains, we have introduced
the anisotropy parameter $\delta$ (i.e. $J_{{}_{/\!/}}^{z} \rightarrow
 J_{{}_{/\!/}} \delta$, and we should set  $\delta=1$ at the isotropic point).
In the limit $y_{1},y_{2} \rightarrow 0$ the exact form of the
above functions are known from the Bethe ansatz  solution
of the chain problem, and, in particular for $\delta=1$
we have $v_{\pm} = \pi/2$, $K_{\pm} =1/2$ \cite{Luther}.
Away from this exactly solvable limit, the expressions (\ref{eq21}) and (\ref{eq22})
should be viewed as  valid
only to lowest order in $\delta$, $y_{1}$ and $y_{2}$, since all
lattice renormalization effects have been neglected in passing  to
the continuum limit. We have  displayed in (\ref{eq20}) only the
relevant (in renormalization group (RG) sense \cite{Schulz})
 operators and have neglected all potentially irrelevant and marginal
ones. The latter contain terms that mix the symmetric and antisymmetric
sectors, as well as combinations of field derivatives and cosines.

The scaling dimensions of the three cosine operators  in (\ref{eq20}) are
$2K_{+}$, $2K_{-}$ and $(2K_{-})^{-1}$ (corresponding to 
$g_{1}$,$g_{2}$ and $g_{3}$, respectively). A cosine operators is
relevant if its scaling dimension is less than two. Thus,
in the limit $y_{1}=y_{2}=0$, all operators have dimension one and
are relevant. For non-zero values of the interchain couplings,
one can easily see that the most relevant operators are
$g_{1}$ and $g_{3}$. These two couplings thus flow to
infinity which signals formation of a gap. This strong
coupling regime corresponds to a non-zero expectation value
of $<\vec{S}_{1}.\vec{S}_{2}>$ \cite{Millis}.  
  Whether a ferromagnetic rung (Haldane) or antiferromagnetic ladder
(Ne\'{e}l) state is realised, depends on the sign of $g_{1}$ and $g_{3}$.
The equations, governing  the RG flow for $g_{1}$ and $g_{3}$  are the usual
Kosterlitz-Thouless equations \cite{Millis}. Thus, 
if initially $g_{1}(0),g_{3}(0) > 0$, then  $g_{1}(l), g_{3}(l)
\rightarrow \infty$, where $l$ is the RG iteration parameter.
This is the N\'{e}el state, characterized by a gap to triplet 
excitations. In the opposite limit $g_{1}(0),g_{3}(0) < 0$
the couplings flow to minus infinity, which is interpreted as
 a ladder with an effective ferromagnetic interchain coupling.       
Thus we conclude that the transition line between   the two states is given
by the equation $g_{1}=g_{3}=y_{1} - 2y_{2} =0$. 

Let us note that as  $y_{1}$ and $y_{2}$  increase 
one should include higher order terms in the operator
product expansion, which leads to (\ref{eq20}). The additional 
terms are still less relevant than the ones in (\ref{eq20}) but
they typically couple the symmetric and antisymmetric sectors.
Thus, the renormalization of $g_{1}$ affects $g_{3}$ and
vice versa. The location of the transition line, however,
is not affected by this coupling.
On the other hand we expect that lattice renormalization
effects, i.e. inclusion of higher order terms in the lattice
spacing, might lead to a change of the shape of the transition line, 
as suggested by numerical simulations (see Fig.2.).

\section{CONCLUSIONS}
We have studied the two-chain antiferromagnetic spin ladder 
with frustrating second neighbour
interactions, using a variety of numerical and analytic
methods. When the interchain nearest neighbour coupling $J_{\perp}$
is dominant the system is in a gapped ``dimerized phase'',
whereas for dominant second neighbour coupling $J_2$ the
system is in a gapped ``Haldane'' phase, which can be
mapped onto an $S=1$ chain. These phases have the same
physical origin within the low energy field theoretic framework
but are distinguished by different behaviour of correlations and
are separated by a first order transition line where there
is a vanishing singlet-singlet gap. The system has a trivial
``valence-bond-solid'' ground state along a line in the
phase diagram which separates two types of classical Ising ground
state. A symmetry of the Hamiltonian allows these states
to be mapped onto each other.

Using  series expansions, about both Ising and dimerized
unperturbed states, we have computed the ground state energy
and dispersion curves for singlet-triplet excitations.
The latter show a qualitative change in form as the second
neighbour interaction changes. Our series in dimer expansions are substantially
longer than in our previous study for $J_2=0$, and we obtain very
accurate estimates for the ground state and excitation gap,
which agree very well with recent Quantum Monte Carlo
results. We have also computed ground state energies and
correlations using exact diagonalizations for $N=16$, 24.
These give a further physical of the nature of the ground
state in different regions of phase diagram.

Finally we present an analytical study, valid for small
$J_{\perp}$ and $J_2$, using the technique of Abelian
bosonization. This gives a picture consistent with the
numerical work.

\acknowledgments
This work forms part of a research project supported by a grant 
from the Australian Research Council. 

\begin{figure}[htb]
\caption{(a) Three different couplings considered: $J_{{}_{/\!/}}$ 
(the dotted lines), $J_{\perp}$ (the dashed lines), 
and $J_2$ (the solid lines). (b) Three different spin orders for the system
at the Ising limit:
the Ne\'{e}l state (N), the ferromagnetic chain state (F),
and the ferromagnetic rung state (R).}
\label{fig:fig1}
\end{figure}

\begin{figure}[htb]
\vspace{9pt}
\caption{Phase diagram in the $(y_1,y_2)$ plane.
The dashed lines show the phase boundary for the
three different spin orders of the system at the Ising limit.
The solid line shows the phase boundary  between 
dimerized phase (right side) and Haldane-type phase (left side)
for the isotropic system. Along this line there is a vanishing singlet-singlet
gap. The ``$\times$'' line ($y_2=1$, $y_1 \geq 1.401484$) is the 
location of the  exact dimer ground states.}
\label{fig:fig2}
\end{figure}

\begin{figure}[htb]
\vspace{9pt}
\caption{The ground-state energy per site $E_0/(NJ_{{}_{/\!/}})$ as 
a function of $y_1=J_{\perp}/J_{{}_{/\!/}}$ for $y_2=0, 0.2, 0.4, 0.6, 0.8, 1$. 
The lines in the large $y_1$ region are the extrapolations of integrated differential
approximants to the dimer series,
the full point symbols connected by a line are estimates from Ising expansions about the
ferromagnetic rung state, and the open point symbols (for $y_2=0$ only) 
are estimates from Ising expansions about the
Ne\'el state. The position of crossing indicate transition 
point with vanishing singlet-singlet gap.
}
\label{fig:fig3}
\end{figure}



\begin{figure}[htb]
\caption{The dispersions $\epsilon_u(k)$ of the spin-triplet excitated states 
of the 2-chain ladder 
with  interchain coupling $y_1=1$ and 
$y_2=0,0.2,0.4,0.6,0.7,0.8,1$. 
The results for $y_2< 0.6$  are estimated from the dimer expansions,
and results for $y_2\ge 0.6$  are estimated from the Ising expansions
about the ferromagnetic rung order.
}
\label{fig:fig4}
\end{figure}

\begin{figure}[htb]
\caption{The dispersions $\epsilon_g(k)$ of the spin-triplet excitated states 
of the 2-chain ladder 
with  interchain coupling $y_1=1$ and 
$y_2=1,0.8,0.7$ (shown in
the figure from  the top to 
the bottom respectively). 
The results are estimated from the Ising expansion 
about the ferromagnetic rung order.
}
\label{fig:fig5}
\end{figure}


\begin{figure}[htb]
\caption{The minimum triplet energy gap 
$\epsilon_u (\pi)/J_{{}_{/\!/}}$
for the system with $J_2=0$
as a function of $J_{\perp}/(J_{\perp}+J_{{}_{/\!/}})$.
The results are estimated from the several different
integrated different approximants to
the dimer series.
}
\label{fig:fig6}
\end{figure}

\begin{figure}[htb]
\caption{
Low lying energies for a system of $N=24$ spins, for fixed
$y_2=0.8$. Solid (dashed) lines represent singlet (triplet) levels
respectively.
}
\label{fig:fig7}
\end{figure}

\begin{figure}[htb]
\caption{
Pair correlations $C_n=4 \langle S_0^z S_n^z \rangle $ in the ground
state for $N=24$ at fixed $y_2=0.8$. Note the discontinuities in 
correlations at the transitions points $y_{1c}\simeq 1.24$ (which 
decreases slightly as the lattice size $N$ increases).
}
\label{fig:fig8}
\end{figure}

\widetext
\begin{table}
\squeezetable
\setdec 0.000000000000
\caption{Series coefficients for the dimer expansion of the
ground-state energy per site $E_0/(NJ_{\perp})$, and the energy gap 
$\epsilon_u (\pi)/J_{\perp}$. Coefficients of
$y_1^{-n}$ are listed for $y_2=0,0.2,0.4,0.6,0.8$. } \label{tab1}
\begin{tabular}{rrrrrr}
 \multicolumn{1}{c}{n} &\multicolumn{1}{c}{$y_2=0$}
&\multicolumn{1}{c}{$y_2=0.2$} &\multicolumn{1}{c}{$y_2=0.4$}
&\multicolumn{1}{c}{$y_2=0.6$} &\multicolumn{1}{c}{$y_2=0.8$}  \\
\tableline
\multicolumn{6}{c}{Ground-state energy $E_0/(NJ_{\perp})$} \\
 0 &\dec $-$3.750000000$\times 10^{-1}$ &\dec $-$3.750000000$\times 10^{-1}$ &\dec $-$3.750000000$\times 10^{-1}$ &\dec $-$3.750000000$\times 10^{-1}$ &\dec $-$3.750000000$\times 10^{-1}$ \\
 1 &\dec  0.000000000 &\dec  0.000000000 &\dec  0.000000000 &\dec  0.000000000 &\dec  0.000000000 \\
 2 &\dec $-$1.875000000$\times 10^{-1}$ &\dec $-$1.200000000$\times 10^{-1}$ &\dec $-$6.750000000$\times 10^{-2}$ &\dec $-$3.000000000$\times 10^{-2}$ &\dec $-$7.500000000$\times 10^{-3}$ \\
 3 &\dec $-$9.375000000$\times 10^{-2}$ &\dec $-$7.200000000$\times 10^{-2}$ &\dec $-$4.725000000$\times 10^{-2}$ &\dec $-$2.400000000$\times 10^{-2}$ &\dec $-$6.750000000$\times 10^{-3}$ \\
 4 &\dec  1.171875000$\times 10^{-2}$ &\dec $-$1.920000000$\times 10^{-2}$ &\dec $-$2.548125000$\times 10^{-2}$ &\dec $-$1.770000000$\times 10^{-2}$ &\dec $-$5.981250000$\times 10^{-3}$ \\
 5 &\dec  8.789062500$\times 10^{-2}$ &\dec  2.880000000$\times 10^{-2}$ &\dec $-$2.953125000$\times 10^{-3}$ &\dec $-$1.080000000$\times 10^{-2}$ &\dec $-$5.146875000$\times 10^{-3}$ \\
 6 &\dec  7.763671875$\times 10^{-2}$ &\dec  4.771200000$\times 10^{-2}$ &\dec  1.317346875$\times 10^{-2}$ &\dec $-$4.452000000$\times 10^{-3}$ &\dec $-$4.292531250$\times 10^{-3}$ \\
 7 &\dec $-$2.682495117$\times 10^{-2}$ &\dec  2.536560000$\times 10^{-2}$ &\dec  1.777094648$\times 10^{-2}$ &\dec  3.552000000$\times 10^{-4}$ &\dec $-$3.463308984$\times 10^{-3}$ \\
 8 &\dec $-$1.381530762$\times 10^{-1}$ &\dec $-$2.345664000$\times 10^{-2}$ &\dec  1.045650727$\times 10^{-2}$ &\dec  3.014910000$\times 10^{-3}$ &\dec $-$2.699781797$\times 10^{-3}$ \\
 9 &\dec $-$1.184420586$\times 10^{-1}$ &\dec $-$5.932056600$\times 10^{-2}$ &\dec $-$3.351507699$\times 10^{-3}$ &\dec  3.491262000$\times 10^{-3}$ &\dec $-$2.033702734$\times 10^{-3}$ \\
10 &\dec  8.023428917$\times 10^{-2}$ &\dec $-$4.401953880$\times 10^{-2}$ &\dec $-$1.476726810$\times 10^{-2}$ &\dec  2.292678800$\times 10^{-3}$ &\dec $-$1.486153249$\times 10^{-3}$ \\
11 &\dec  2.926602935$\times 10^{-1}$ &\dec  2.352668756$\times 10^{-2}$ &\dec $-$1.616860920$\times 10^{-2}$ &\dec  2.872863433$\times 10^{-4}$ &\dec $-$1.066983590$\times 10^{-3}$ \\
12 &\dec  2.174702423$\times 10^{-1}$ &\dec  9.158584993$\times 10^{-2}$ &\dec $-$5.942062903$\times 10^{-3}$ &\dec $-$1.585372406$\times 10^{-3}$ &\dec $-$7.754224757$\times 10^{-4}$ \\
13 &\dec $-$2.513295675$\times 10^{-1}$ &\dec  8.611973765$\times 10^{-2}$ &\dec  9.811790575$\times 10^{-3}$ &\dec $-$2.607764454$\times 10^{-3}$ &\dec $-$6.016035581$\times 10^{-4}$ \\
14 &\dec $-$7.079677824$\times 10^{-1}$ &\dec $-$2.062191934$\times 10^{-2}$ &\dec  2.034441425$\times 10^{-2}$ &\dec $-$2.487025970$\times 10^{-3}$ &\dec $-$5.287392850$\times 10^{-4}$ \\
15 &\dec $-$4.220720468$\times 10^{-1}$ &\dec $-$1.567438679$\times 10^{-1}$ &\dec  1.702026317$\times 10^{-2}$ &\dec $-$1.402179099$\times 10^{-3}$ &\dec $-$5.356521930$\times 10^{-4}$ \\
16 &\dec  8.048506089$\times 10^{-1}$ &\dec $-$1.794371390$\times 10^{-1}$ &\dec $-$1.145012657$\times 10^{-4}$ &\dec  1.136830590$\times 10^{-4}$ &\dec $-$5.993625741$\times 10^{-4}$ \\
17 &\dec  1.836906814 &\dec  1.631560655$\times 10^{-4}$ &\dec $-$2.064904725$\times 10^{-2}$ &\dec  1.407604936$\times 10^{-3}$ &\dec $-$6.974813639$\times 10^{-4}$ \\
18 &\dec  7.873287228$\times 10^{-1}$ &\dec  2.827706741$\times 10^{-1}$ &\dec $-$2.936191639$\times 10^{-2}$ &\dec  1.968009179$\times 10^{-3}$ &\dec $-$8.101904773$\times 10^{-4}$ \\
19 &\dec $-$2.609038806 &\dec  3.876215459$\times 10^{-1}$ &\dec $-$1.667240377$\times 10^{-2}$ &\dec  1.615756127$\times 10^{-3}$ &\dec $-$9.216722624$\times 10^{-4}$ \\
20 &\dec $-$4.952285110 &\dec  7.857208468$\times 10^{-2}$ &\dec  1.260776202$\times 10^{-2}$ &\dec  5.560918471$\times 10^{-4}$ &\dec $-$1.020922924$\times 10^{-3}$ \\
21 &\dec $-$1.190569396 &\dec $-$5.222477673$\times 10^{-1}$ &\dec  3.933197151$\times 10^{-2}$ &\dec $-$7.233155669$\times 10^{-4}$ &\dec $-$1.101950155$\times 10^{-3}$ \\
22 &\dec  8.511630761 &\dec $-$8.553338679$\times 10^{-1}$ &\dec  4.125471073$\times 10^{-2}$ &\dec $-$1.663509824$\times 10^{-3}$ &\dec $-$1.163436744$\times 10^{-3}$ \\
23 &\dec  1.360832756$\times 10^{1}$ &\dec $-$3.290829443$\times 10^{-1}$ &\dec  9.852975277$\times 10^{-3}$ &\dec $-$1.867815929$\times 10^{-3}$ &\dec $-$1.207975395$\times 10^{-3}$ \\
\tableline
\multicolumn{6}{c}{Energy gap $\epsilon_u (\pi)/J_{\perp}$} \\
 0 &\dec   1.000000000 &\dec   1.000000000 &\dec   1.000000000 &\dec   1.000000000 &\dec   1.000000000 \\
 1 &\dec $-$1.000000000 &\dec $-$8.000000000$\times 10^{-1}$ &\dec $-$6.000000000$\times 10^{-1}$ &\dec $-$4.000000000$\times 10^{-1}$ &\dec $-$2.000000000$\times 10^{-1}$ \\
 2 &\dec   5.000000000$\times 10^{-1}$ &\dec   3.200000000$\times 10^{-1}$ &\dec   1.800000000$\times 10^{-1}$ &\dec   8.000000000$\times 10^{-2}$ &\dec   2.000000000$\times 10^{-2}$ \\
 3 &\dec   2.500000000$\times 10^{-1}$ &\dec   1.600000000$\times 10^{-1}$ &\dec   9.000000000$\times 10^{-2}$ &\dec   4.000000000$\times 10^{-2}$ &\dec   1.000000000$\times 10^{-2}$ \\
 4 &\dec $-$1.250000000$\times 10^{-1}$ &\dec $-$6.080000000$\times 10^{-2}$ &\dec $-$3.780000000$\times 10^{-2}$ &\dec $-$2.480000000$\times 10^{-2}$ &\dec $-$9.800000000$\times 10^{-3}$ \\
 5 &\dec $-$2.734375000$\times 10^{-1}$ &\dec $-$1.596800000$\times 10^{-1}$ &\dec $-$1.098225000$\times 10^{-1}$ &\dec $-$7.624000000$\times 10^{-2}$ &\dec $-$3.200750000$\times 10^{-2}$ \\
 6 &\dec $-$1.533203125$\times 10^{-1}$ &\dec $-$1.282880000$\times 10^{-1}$ &\dec $-$1.210798125$\times 10^{-1}$ &\dec $-$1.114720000$\times 10^{-1}$ &\dec $-$5.737081250$\times 10^{-2}$ \\
 7 &\dec   2.456054688$\times 10^{-1}$ &\dec   5.984960000$\times 10^{-2}$ &\dec $-$4.939160625$\times 10^{-2}$ &\dec $-$1.189328000$\times 10^{-1}$ &\dec $-$8.546363125$\times 10^{-2}$ \\
 8 &\dec   4.813385010$\times 10^{-1}$ &\dec   2.335137600$\times 10^{-1}$ &\dec   5.900339848$\times 10^{-2}$ &\dec $-$9.919344000$\times 10^{-2}$ &\dec $-$1.162613828$\times 10^{-1}$ \\
 9 &\dec   1.322269440$\times 10^{-1}$ &\dec   2.147366880$\times 10^{-1}$ &\dec   1.432105375$\times 10^{-1}$ &\dec $-$5.701525600$\times 10^{-2}$ &\dec $-$1.496337233$\times 10^{-1}$ \\
10 &\dec $-$6.962262789$\times 10^{-1}$ &\dec $-$7.397652400$\times 10^{-2}$ &\dec   1.363951349$\times 10^{-1}$ &\dec $-$5.152148667$\times 10^{-3}$ &\dec $-$1.855936645$\times 10^{-1}$ \\
11 &\dec $-$1.056785534 &\dec $-$4.262824117$\times 10^{-1}$ &\dec   2.673760586$\times 10^{-2}$ &\dec   4.221120509$\times 10^{-2}$ &\dec $-$2.241092352$\times 10^{-1}$ \\
12 &\dec   5.050360756$\times 10^{-2}$ &\dec $-$4.592008720$\times 10^{-1}$ &\dec $-$1.317976096$\times 10^{-1}$ &\dec   7.169697802$\times 10^{-2}$ &\dec $-$2.651208173$\times 10^{-1}$ \\
13 &\dec   2.122963428 &\dec   6.458482125$\times 10^{-2}$ &\dec $-$2.320399129$\times 10^{-1}$ &\dec   7.586016636$\times 10^{-2}$ &\dec $-$3.084074357$\times 10^{-1}$ \\
\end{tabular}
\end{table}

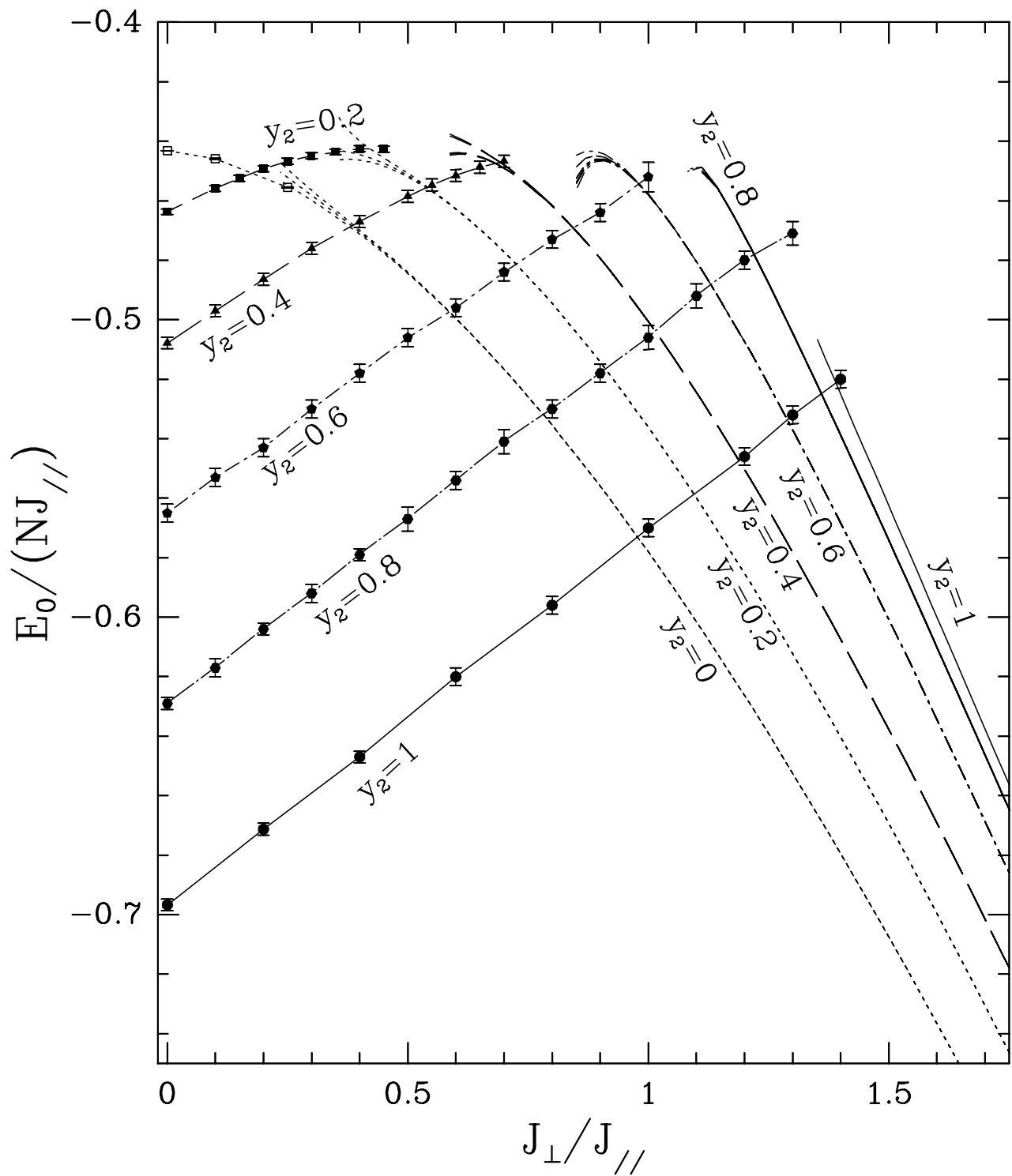
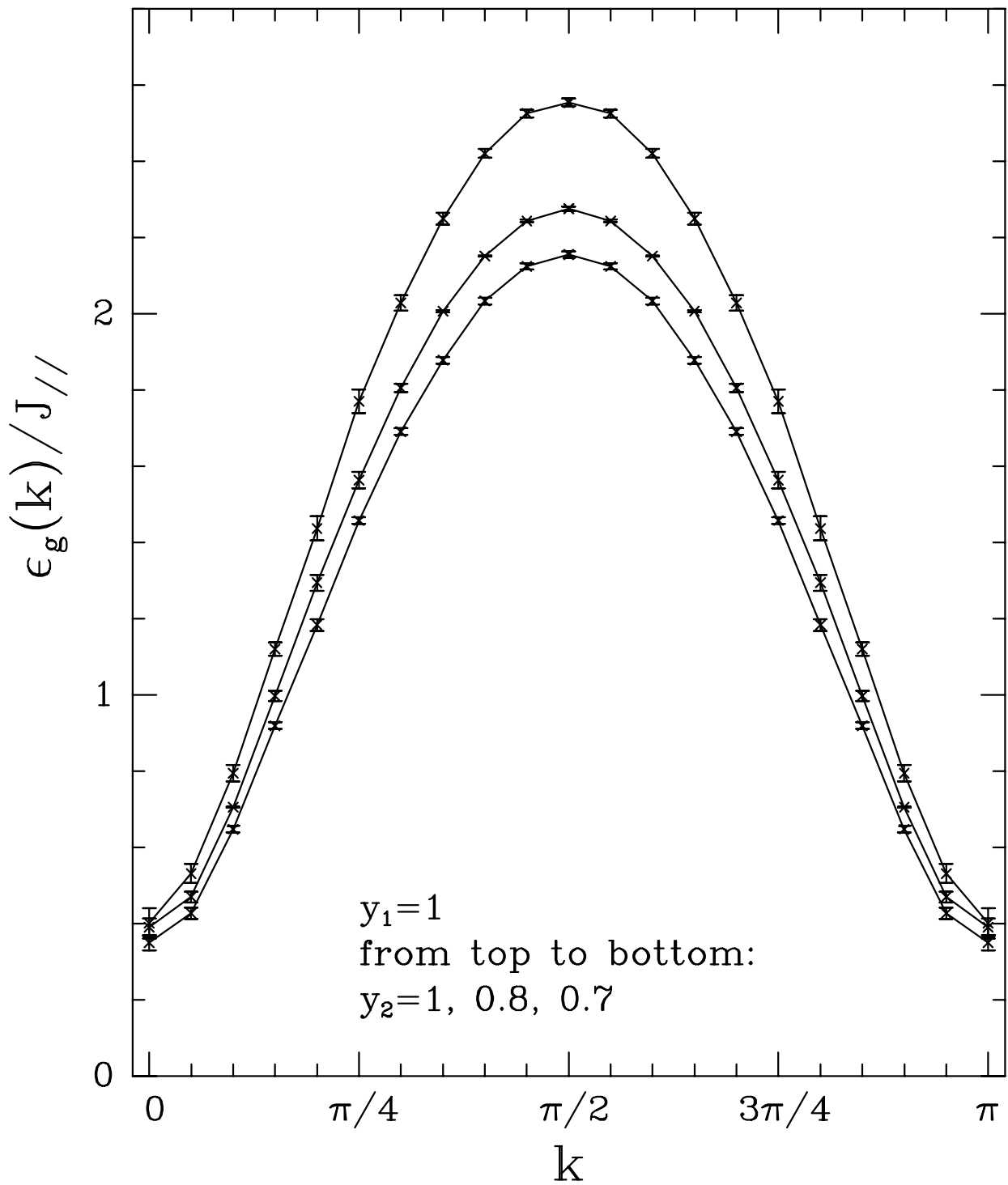
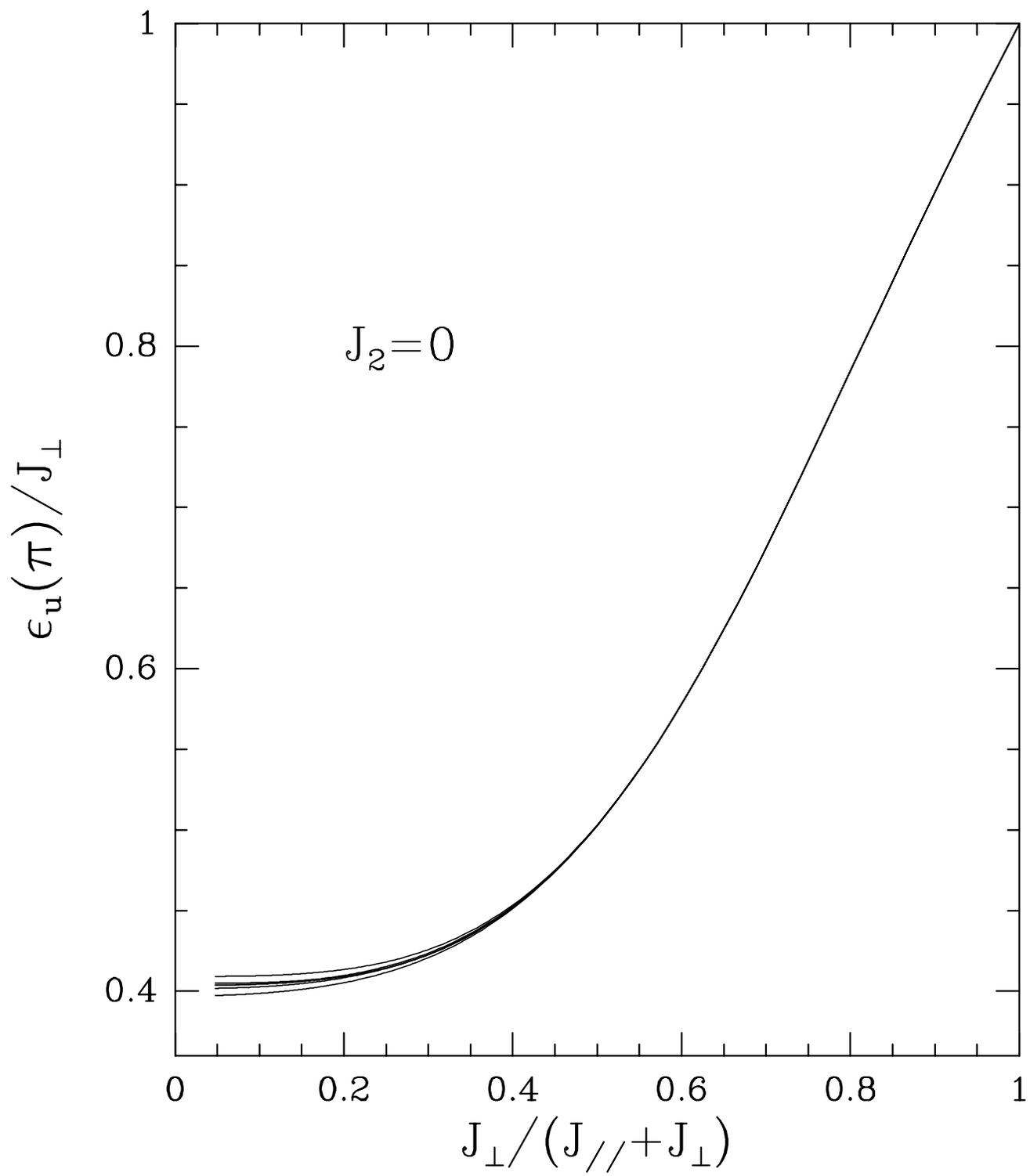
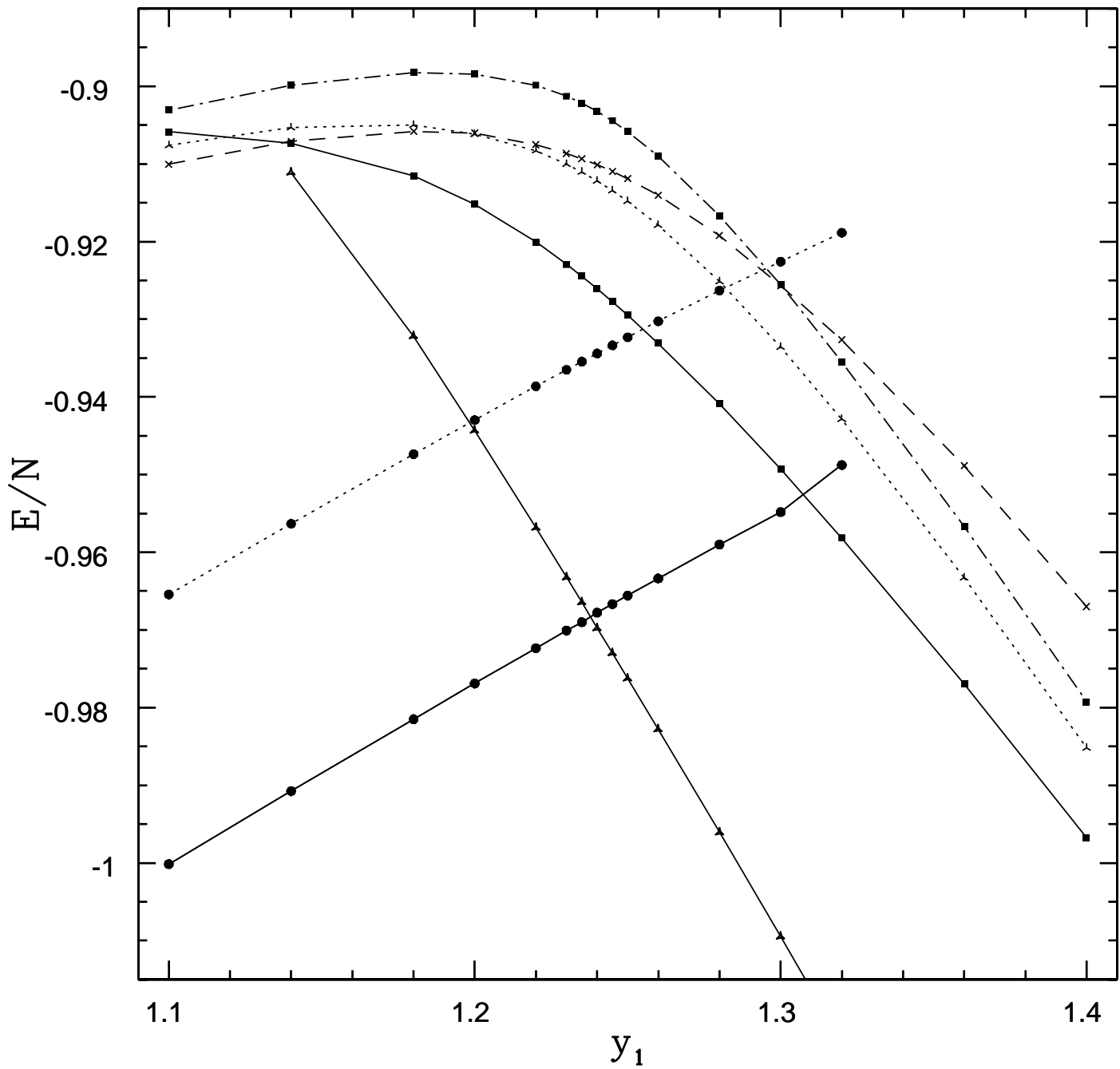
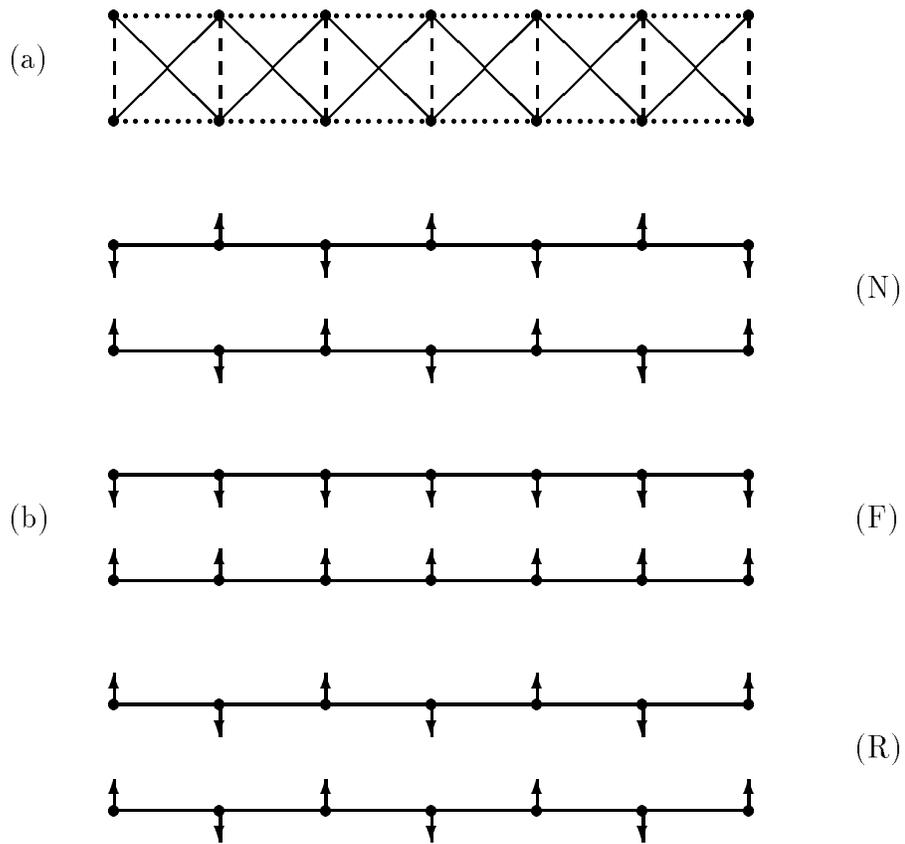
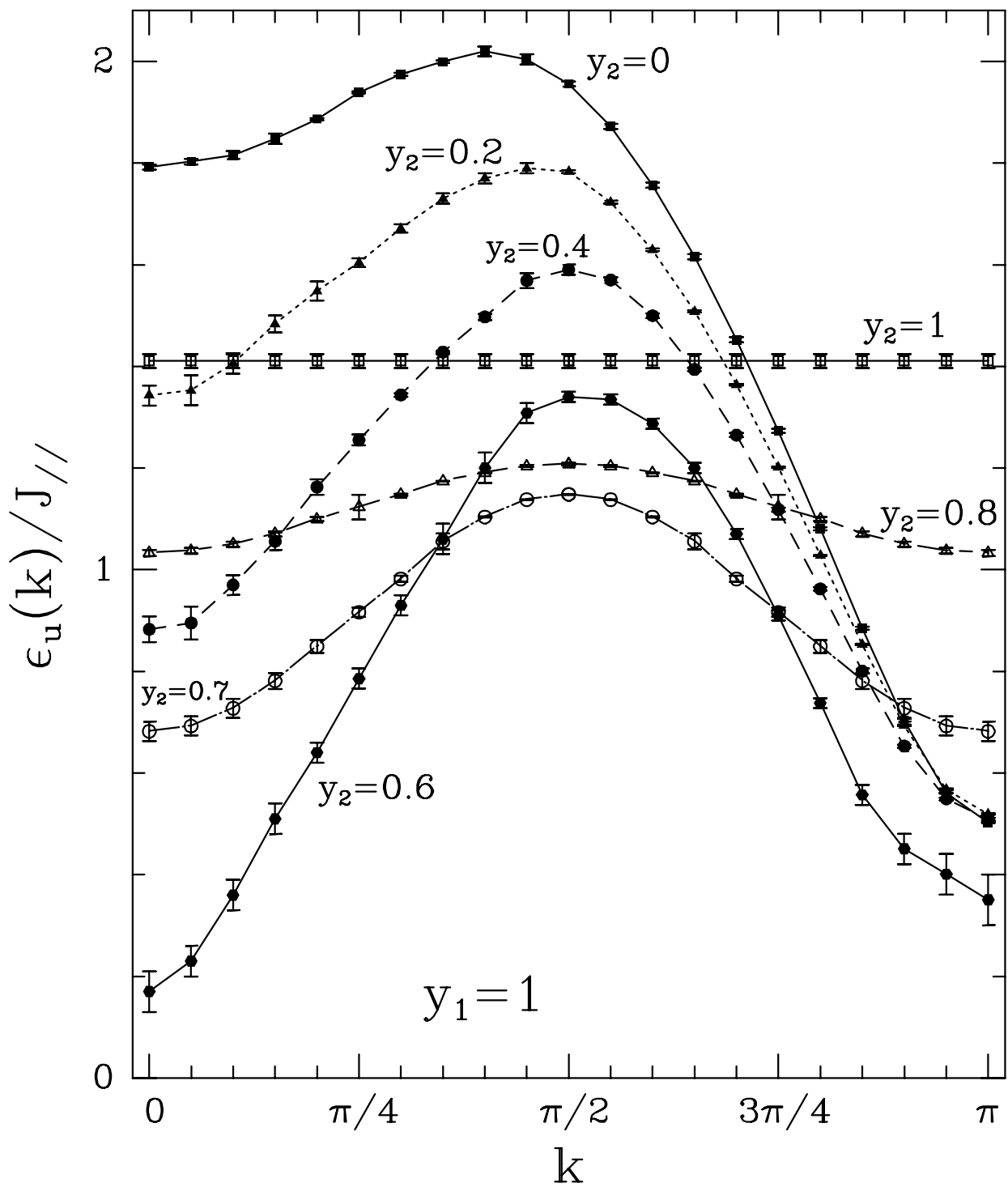
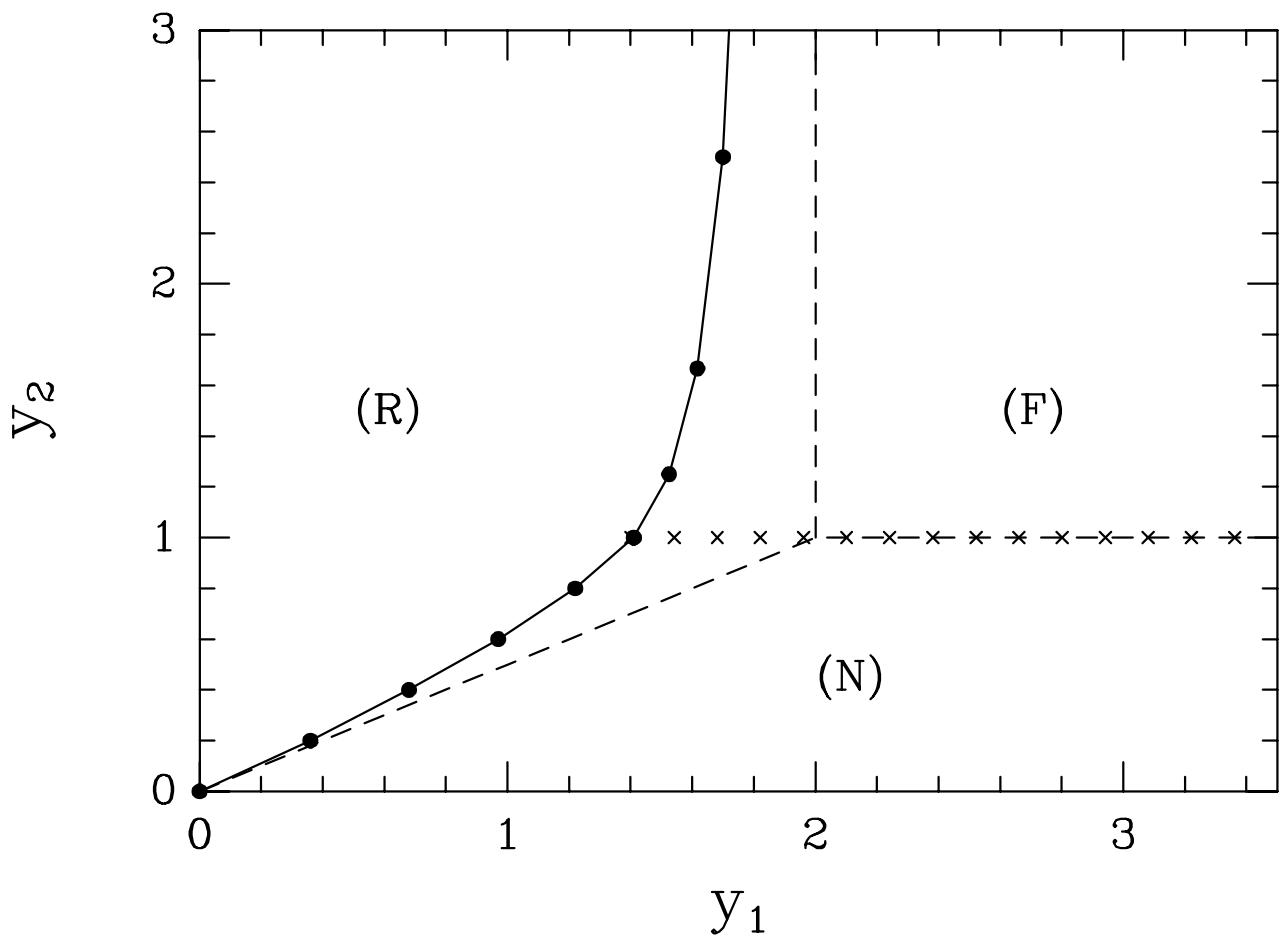
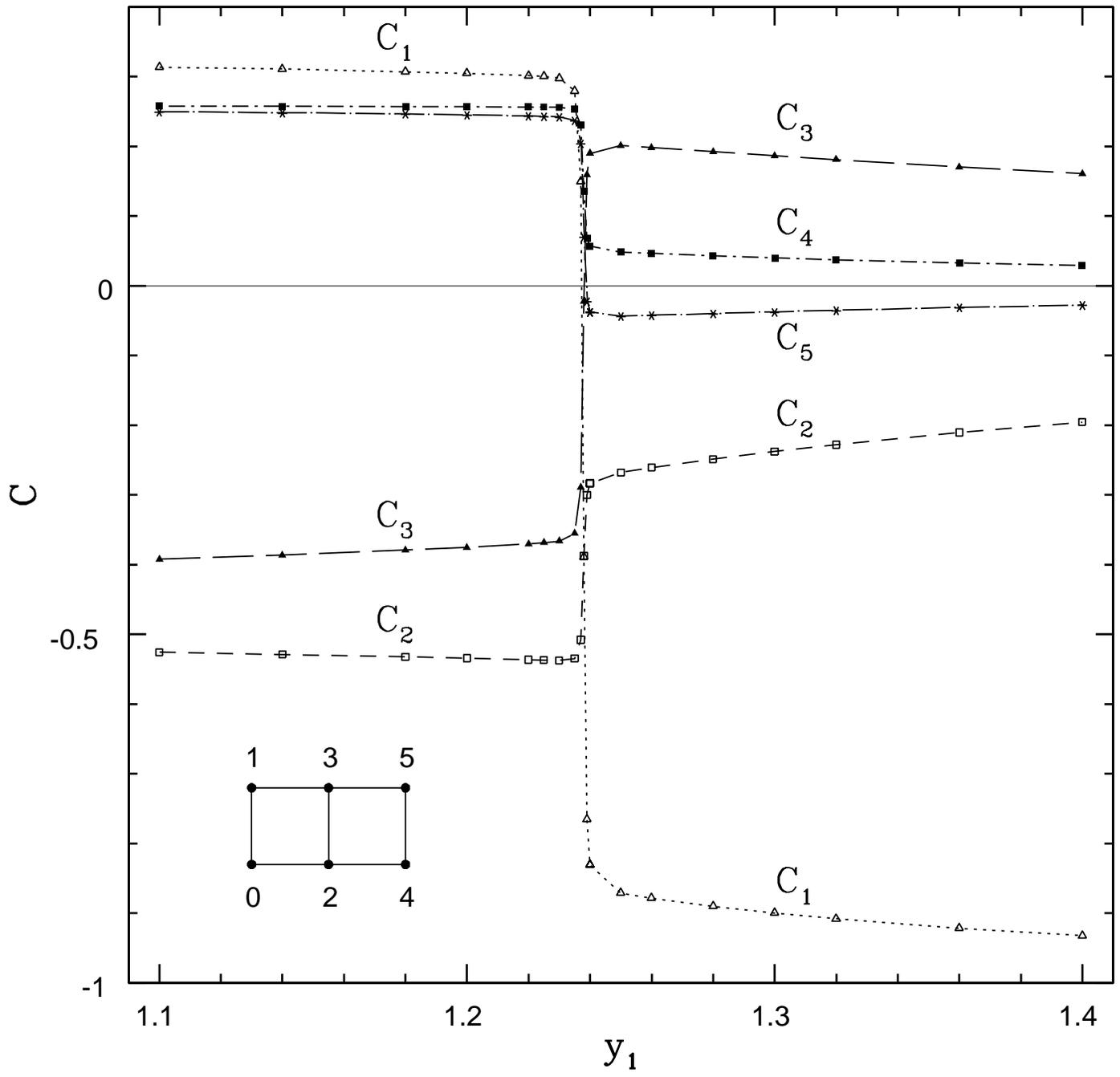
\end{document}